# Anomalous Fraunhofer patterns in $Cd_3As_2$ Josephson Junctions

Rak-Hee Kim[1], Yeongmin Jang[1], Bob M. Wang[2], Dong Yu[2], Yong-Joo Doh[1,§]

[1]Department of Physics and Photon Science, Gwangju Institute of Science and Technology (GIST), Gwangju, 61005, Republic of Korea
[2]Department of Physics, University of California, Davis, CA, 95616, USA

## Abstract

Majorana zero modes (MZMs) in topological superconductors are promising for quantum computing, yet their unambiguous detection remains challenging. We fabricated Josephson junctions (JJs) using $Cd_3As_2$ Dirac semimetal nanoribbons with NbTi superconducting electrodes to investigate topological supercurrents through Fraunhofer pattern analysis. The JJs exhibited excellent quality with high transparency ($\tau = 0.77$) and large induced superconducting gap ($\Delta = 1.10$ meV), confirmed by multiple Andreev reflection features. While node lifting at the third minimum of the Fraunhofer pattern was observed as a predicted signature of $4\pi$-periodic topological supercurrents, our theoretical analysis demonstrates that asymmetric supercurrent distributions can reproduce this behavior without invoking MZMs. These findings reveal that anomalous Fraunhofer patterns alone cannot reliably confirm topological superconductivity, necessitating complementary experimental approaches for conclusive Majorana detection.

# 1. Introduction

Majorana zero modes (MZMs), topologically protected quasiparticle states in topological superconductors, are regarded as promising building blocks for fault-tolerant topological quantum computing [1, 2]. Experimental efforts to detect MZMs have primarily focused on the III-V semiconductor heterostructures with strong spin-orbit coupling [3, 4] and topological insulators (TIs) [5, 6] in proximity to conventional $s$-wave superconductors. Signatures including zero-bias conductance peak (ZBCP) in tunneling junctions [3, 4, 7] and missing odd-integer Shapiro steps, known as the fractional Josephson effect, in topological Josephson junctions (JJs) [5, 6] have been interpreted as evidence for MZMs and topological supercurrent, respectively. However, in recent years, similar ZBCP and fractional Josephson effect features have also been reported in topologically trivial systems [8, 9], raising controversy over the unambiguous identification of MZMs.

Recently, Fraunhofer pattern analysis has been proposed as a criterion to verify $4\pi$-periodic topological supercurrents [10, 11]. This approach utilizes the characteristic modulation of the critical current in a JJ by a perpendicular magnetic field to the substrate. In conventional JJs with a $2\pi$-periodic current-phase relation (CPR), the Fraunhofer pattern exhibits nodes spaced by the magnetic flux quantum $\Phi_0 = h/2e$, where $h$ is Planck's constant and $e$ is the elementary charge [12]. By contrast, topological JJs are predicted to exhibit nodes separated by $2\Phi_0$, a hallmark of the $4\pi$-periodic CPR. When conventional and topological supercurrents coexist, the Fraunhofer pattern is expected to show node lifting at odd multiples of $\Phi_0$ [10].

Dirac semimetal (DSM) is a three-dimensional topological material characterized by a linear band dispersion and four-fold degenerate Dirac points in the bulk, protected by the combined symmetries of time-reversal, inversion, and crystalline structure [13]. Theoretically, DSMs are expected to provide a promising platform for realizing topological superconductivity through the superconducting proximity effects [14, 15]. As a representative DSM, $Cd_3As_2$ nanowires (NWs) and nanoribbons (NRs) have exhibited Aharonov-Bohm oscillations arising from surface states [16], while $Cd_3As_2$ flakes have demonstrated supercurrent transport through hinge states, a hallmark of higher-order topological semimetals [17]. Furthermore, ZBCP [18] and the absence of the first Shapiro step [19] have been reported in superconducting junctions of $Cd_3As_2$. However, other studies have found no signatures of topological Josephson effects in DSM-based JJs [20, 21]. Therefore, experimental evidence of topological supercurrent and MZMs in DSMs remains inconclusive.

In this study, we fabricated JJs based on $Cd_3As_2$ NRs and performed low-temperature transport measurements under varying temperature and magnetic field. The JJs exhibited clear supercurrent branches and multiple differential-conductance peaks, indicating highly transparent contacts are formed. When a magnetic field is applied perpendicular to the substrate, the critical current oscillates with the field and shows node lifting at the third minimum of the Fraunhofer pattern. While such anomalous Fraunhofer patterns are often attributed to topological supercurrents mediated by MZMs, our numerical analysis suggests that the node lifting can also originate from non-topological factors, such as an asymmetric distribution of the critical current within the JJ. These results demonstrate that anomalous Fraunhofer patterns alone cannot provide a sufficiently reliable criterion for identifying topological supercurrent.

## 2. Experimental method

$Cd_3As_2$ NWs and NRs were grown on Si substrates using chemical vapor deposition method [22]. Figure 1a shows transmission electron microscope (TEM) image of $Cd_3As_2$ NW, revealing an interlayer spacing of 0.73 nm along the axial growth direction of [112], consistent with previous reports [22, 23]. For device fabrication, NWs were manually transferred using a tungsten tip onto an *n*-type Si substrate with a 300-nm-thick $SiO_2$ layer. Electrode patterns were defined by electron beam lithography, followed by oxygen plasma treatment to remove residual electron beam resist. Prior to metallization, the NW surface was etched by Ar ion milling (etching rate = 6 nm/min) inside an electron beam evaporator chamber to remove the native oxide. Without breaking vacuum in the chamber, Ti/Au (10 nm/80 nm) electrodes were deposited by electron beam evaporation for normal-metal contacts, while NbTi (56 nm) electrodes were deposited by magnetron sputtering using a 99.9% pure NbTi (50/50 at.%) target at a rate of 5.8 nm/min. The resist was lifted off in acetone after the metal deposition. Figure 1b presents a scanning electron microscopy (SEM) image of a $Cd_3As_2$ NW device contacted with Ti/Au electrodes. Transport measurements were carried out in a closed-cycle $^4$He cryostat system (Seongwoo Instruments Inc.) with a base temperature of 2.7 K. To reduce electrical noise, low-pass RC filters and π filters were connected in series with the measurement lines [24].

## 3. Results and discussion

Resistance ($R$) as a function of gate voltage ($V_g$) measured at $T = 2.7$ K is presented in Fig. 1c for $Cd_3As_2$ NW devices with different channel lengths. The NW diameter is $d = 79$ nm, as shown in the inset of Fig. 1d, while the channel lengths are $L = 1.0$ μm (**D1**), 2.0 μm (**D2**), and 3.0 μm (**D3**). A back gate electrode was used to apply the gate voltage during the $R(V_g)$ measurements. All devices exhibit ambipolar transport behavior with maximum resistance at the Dirac point ($V_{DP}$), which occurs at $V_{DP} = 1.54$ V (**D1**), $-0.34$ V (**D2**), and 0.81 V (**D3**). From the conductance ($G$) vs. ($V_g - V_{DP}$) curves shown in Fig. 1d, we extract the derivative transconductance $dG/dV_g = 1.5 \times 10^{-5}$ S/V, a carrier mobility $\mu = 1.1 \times 10^4$ cm$^2$/Vs, a carrier concentration $n = 3.1 \times 10^{16}$ cm$^{-3}$ at $V_g = 0$ V, and an elastic mean free path $l_e = 72$ nm, using the gate capacitance $C_g = 1.15 \times 10^{-16}$ F [25]. The calculations employ effective mass $m^* = 0.053\, m_e$ and Fermi velocity $v_F = 2.12 \times 10^5$ m/s, where $m_e$ is the free electron mass [26]. These transport characteristics are consistent with previous reports, which show $n$ in the range of $10^{16}$–$10^{17}$ cm$^{-3}$ and mobilities of $10^3$–$10^4$ cm$^2$/Vs [26-29].

Figure 2a shows an SEM image of two $Cd_3As_2$ NR JJs, denoted **J1** and **J2**. The NR has a thickness of $t = 44$ nm and a width of $w = 317$ nm, with junction channel lengths of $L = 125$ nm (**J1**) and 268 nm (**J2**). Figure 2b presents the current–voltage ($I$–$V$) characteristics measured at 2.7 K, which display clear supercurrent branches arising from the superconducting proximity effect [12]. The critical currents are $I_c = 1.74$ μA for **J1** and 0.23 μA for **J2**, with corresponding normal-state resistances of $R_n = 155\ \Omega$ and 254 $\Omega$, respectively. No hysteresis was observed in the $I$–$V$ curves.

The critical current decreases monotonically with increasing temperature, as shown in Fig. 2c, and vanishes at 6.9 K (5.1 K) for device **J1** (**J2**), respectively (Fig. 2d). In the long and diffusive junction limit, the temperature dependence of the critical current is described by the relation: $eI_c(T)R_n = aE_{Th}[1 - b\exp(-aE_{Th}/3.2k_BT)]$, where $E_{Th}$ is the Thouless energy, $k_B$ is the Boltzmann constant, and $a$, $b$ are fitting parameters [30]. The Thouless energy is calculated as $E_{Th} = \hbar D/L^2$, yielding $E_{Th}$ = 185 μeV for **J1** and 40 μeV for **J2**, with $\hbar$ the reduced Planck constant and $D = v_F l_e/3$ the diffusion constant. The solid lines in Fig. 2d represent the fitting results using parameters $a$ = 3.50 (6.54) and $b$ = 1.45 (1.23) for **J1** (**J2**), showing good agreement with the experimental data. The superconducting coherence length is estimated as $\xi = \sqrt{\hbar D/\Delta}$, giving $\xi$ = 52 nm, where $\Delta(0)$ = 1.10 meV is the superconducting gap at zero temperature (presented later). Since the junction length $L \gg \xi$ and the electric transport is diffusive ($l_e < L$), both J1 and J2 meet the criteria of the long and diffusive junction limit.

Figure 3a shows the differential conductance (d$I$/d$V$) of device **J1** as a function of bias voltage at 2.7 K. Several distinct d$I$/d$V$ peaks, highlighted by arrows, are observed and evolve systematically with temperature (Fig. 3b). These features are attributed to multiple Andreev reflections (MARs) occurring at the interface between the NR and the superconducting electrodes. In a normal metal-superconductor (N-S) junction with highly transparent contact, when an electron with energy below the superconducting gap is incident on the interface, it is retro-reflected as a hole while a Cooper pair is transferred into the superconductor. This process, known as Andreev reflection [31], enhances the conductance of the junction. For a JJ with an S-N-S structure, repeated Andreev reflections of holes and electrons at both interfaces give rise to MAR, leading to d$I$/d$V$ peaks at discrete voltages $V_n = 2\Delta/ne$, where $n$ is an integer [32, 33]. Consequently, the peaks labeled in Fig. 3a can be identified as MAR features, providing a direct spectroscopic measure of the superconducting gap $\Delta$ as $V_2 = \Delta/e$ and $V_3 = 2\Delta/3e$. Figure 3c displays the temperature dependence of $\Delta(T)$, extracted from the peak positions of $V_2$ and $V_3$. The solid line represents the Bardeen-Cooper-Schrieffer (BCS) theoretical expectation [12], showing good agreement with the experimental data. From the fit, we obtain $\Delta(0)$ = 1.10 meV and a superconducting transition temperature $T_c$ = 8.7 K.

The I–V characteristics of **J1** and **J2**, measured over a wide voltage range at 2.7 K, are displayed in Fig. 3d. Due to conductance enhancement from MAR, an excess current ($I_{exc}$) is clearly observed, defined as the intercept on the current axis obtained from the linear extrapolation of the high-bias voltage regime ($V > 2\Delta/e$). The extracted values are $I_{exc}$ = 8.45 μA for **J1** and 4.17 μA for **J2**. Since the junction transparency $\tau$ is proportional to $eI_{exc}R_n/\Delta$ [34], we obtain $\tau = 0.77$ for **J1**, and 0.71 for **J2**. These values are close to the ideal limit of $\tau = 1.0$, demonstrating that highly transparent contacts are formed in our JJs. The obtained transparencies are comparable to the best results reported in JJs based on $Cd_3As_2$ nanoplates [16, 17] and Sb-doped $Bi_2Se_3$ topological insulator NRs [35]. Using the extracted $\tau$ and $\Delta$, the magnitude of the 4π-periodic topological supercurrent for **J1** can be estimated as [36] $I_{c,4\pi} = \sqrt{\tau}e\Delta/\hbar = 236$ nA.

The magnetic field ($B$) dependence of the critical current of **J1** at 2.7 K is presented in Fig. 4a, with the field applied perpendicular to the substrate. $I_{c+}$ and $I_{c-}$ denote the critical currents for positive and negative bias polarities, respectively. The two $I_c$'s are symmetric with each other and exhibit quasi-periodic oscillations with the magnetic field. For conventional (or non-topological) JJs, the field dependence of $I_c$ is expected to follow the well-known Fraunhofer pattern relation [12]: $I_c(\Phi) = I_c(0)|\sin(\pi\Phi/\Phi_0)/(\pi\Phi/\Phi_0)|$, where $\Phi_0 = h/2e$ is the magnetic flux quantum and $\Phi = BL_{eff}w$ is the flux through the junction area. Here, $L_{eff} = L + 2\lambda_L$ represents the effective junction length, which includes both the physical junction length $L$ and the London penetration depth $\lambda_L$ inside the superconducting electrodes. Using $L_{eff} = 258$ nm and $\lambda_L = 67$ nm, the Fraunhofer-type calculation (solid line) shows good agreement with the experimental data of $I_{c+}$, except for the finite $\Delta I_c = I_{c+} - I_{c-} = 44$ nA observed at the third node of the $I_c(B)$ oscillation data (see the inset). Node lifting at the odd-integer minima of $I_c(B)$ oscillations is often attributed to a 4π-periodic topological supercurrent induced by MZMs in topological JJs [10]. However, to assess the reliability of this criterion, it is necessary to also consider non-topological scenarios that could account for the anomalous $I_c(B)$ behavior.

First, we consider the geometric effect of the NR used in this experiment. Owing to its finite thickness ($t \approx 44$ nm), supercurrents may also flow through junctions formed along the NR sidewalls, leading to an enhancement of $J_c(x)$ near the edges. Figure 4b shows the calculated $I_c(B)/I_c(0)$ patterns for various edge-enhanced $J_c(x)$ profiles in the inset. As the edge supercurrent contribution increases, the magnitude of $I_c(B)$ lobes becomes larger, resembling those of a conventional superconducting quantum interference device (SQUID) [12]. However, in all simulated $J_c(x)$ profiles, the nodal points of $I_c(B)$ remain fully closed, in contrast to our experimental observations. Therefore, the edge-enhanced supercurrent scenario alone cannot account for the node lifting observed in this work.

Second, we analyze $I_c(B)$ pattern to extract the spatial distribution of the critical current density $J_c(x)$. By applying the inverse Fourier transform to the $I_c(B)$ data, $J_c(x)$ can be obtained using the Dynes-Fulton method [37]. Figure 4c shows $J_c(x)$ extracted from the $I_c(B)$ data of **J1**, revealing a slight asymmetry with respect to the $x = 0$ axis. Using a simplified model of non-uniform $J_c(x)$, illustrated in the inset of Fig. 4d, we calculated the critical current [38]

$$I_c = \left| \int_{-\infty}^{\infty} J_c(x) e^{i\beta x} dx \right| \qquad (1)$$

with $\beta = 2\pi B L_{\text{eff}}/\Phi_0$ and $L_{\text{eff}} = 258$ nm. The calculated result (solid line, Fig. 4d) reproduces the experimental data (symbols), including the selective lifting of the third node. Although the asymmetric component, $\int_{-\infty}^{\infty} \{|J_c(x) - J_c(-x)|/2\} dx$, corresponds to only ~ 2% of the zero-field critical current, the resulting $I_c(B)$ oscillations deviate from the conventional Fraunhofer pattern and exhibit anomalous nodal lifting. Since the critical current in the long-junction limit decays exponentially with channel length, $I_c \sim \exp(-L/\xi)$ [39], variations in channel length or junction transparency can induce asymmetries in $J_c(x)$ and modify the Fraunhofer pattern. Therefore, observation of node lifting in $I_c(B)$ alone is not a sufficiently reliable criterion for demonstrating the existence of MZMs in topological JJs.

## 4. Conclusion

We fabricated highly transparent Josephson junctions using $Cd_3As_2$ nanoribbons and NbTi superconducting electrodes to investigate odd-node lifting in the Fraunhofer pattern as a potential signature of topological supercurrent. The junctions exhibit robust supercurrents and pronounced differential-conductance peaks arising from multiple Andreev reflections, demonstrating excellent junction quality with high transparency ($\tau = 0.77$) and a large induced superconducting gap ($\Delta(0) = 1.10$ meV). While node lifting was observed at the third node of the Fraunhofer pattern, our analysis indicates that this feature can be explained by non-topological origins, specifically asymmetric supercurrent distributions within the junction, rather than by Majorana zero modes. Our observations highlight the challenges in identifying topological supercurrent based solely on anomalous Fraunhofer pattern and the need for complementary experimental approaches to provide conclusive evidence for Majorana zero modes in topological Josephson junctions.

## Author contributions

R.H.K. fabricated the devices and performed low-temperature measurements. R.H.K. and Y.J. analyzed the data and carried out the calculations. B.M.W. and D.Y. synthesized the samples and performed structural characterization. Y.J.D. designed and supervised the research project. The manuscript was written by R.H.K., Y.J., and Y.J.D. with input from all authors.

## Declaration of competing interest

The authors declare that they have no known competing financial interests or personal relationships that could have appeared to influence the work reported in this paper.

## Acknowledgement

This study was supported by the NRF of Korea through the Basic Science Research Program (RS-2018-NR030955, RS-2023-00207732, RS-2025-02317602), the ITRC program (IITP-2025-RS-2022-00164799) funded by the Ministry of Science and ICT, and the GIST Research Project grant funded by the GIST in 2025.

## Data availability

All data in this study are available on request.

**Figure captions**

**Fig. 1** Structural characterization and electrical transport of $Cd_3As_2$ NW devices. (a) TEM image of $Cd_3As_2$ NW, showing an interplanar spacing of 0.73 nm along the [112] axial-growth direction (indicated by the arrow). (b) Representative SEM image of a $Cd_3As_2$ NW with Ti/Au electrodes. (c) Resistance $R$ as a function of gate voltage $V_g$ for NW devices with different channel lengths: $L$ = 1.0 μm (D1), 2.0 μm (D2) and 3.0 μm (D3). (d) Conductance $G$ vs. $V_g - V_{DP}$, extracted from (c), where $V_{DP}$ is the gate voltage at the Dirac point. Inset: AFM height profile of the NW

**Fig. 2** Transport characteristics of $Cd_3As_2$ NR JJs. (a) SEM image of the fabricated NbTi-$Cd_3As_2$ NR-NbTi JJs, labeled J1 and J2. The bias current was applied from I+ to I–, and the voltage drop was measured between V+ and V–. (b) Current-voltage (*I-V*) characteristics of J1 and J2 measured at $T$ = 2.7 K. The critical current $I_c$ is indicated for J1. Inset: AFM height profile of the NR. (c) Temperature-dependent *I-V* curves of J1. (d) Temperature dependent critical current $I_c(T)$ for J1 (squares) and J2 (circles). The solid lines represent theoretical fits for the long and diffusive junction limit

**Fig. 3** Quantum electronic transport properties of $Cd_3As_2$ NR JJs. (a) Differential conductance d$I$/d$V$ of J1 as a function of voltage at $T$ = 2.7 K. Arrows mark the conductance peaks that occur at voltages $V_n$ indexed by integer $n$. (b) Temperature dependence of the d$I$/d$V(V)$ characteristics of J1. (c) Temperature dependence of the superconducting gap energy $\Delta$, extracted from the peak voltages $V_2$ (circles) and $V_3$ (squares) of J1. The solid line shows the BCS prediction for $\Delta(T)$. (d) *I-V* characteristics of J1 and J2 over a wide bias range (solid lines). Dashed lines denote the extrapolation used to determine the excess currents $I_{exc}$

**Fig. 4** Magnetic field dependence of the critical current. (a) Critical current $I_c$ as a function of magnetic field $B$. $I_{c+}$ (red) and $I_{c-}$ (blue) correspond to the critical currents for positive and negative current bias polarities, respectively, when the bias is swept from negative to positive. The solid line represents a fit to the non-topological Fraunhofer pattern. Inset: Enlarged view of the nodal regions. (b) Calculated $I_c(B)/I_c(0)$ curves obtained from the $J_c(x)$ profiles shown in the inset. Parameters are set to $w$ = 400 nm and $L_{eff}$ = 200 nm. Inset: $J_c(x)$ profiles representing uniform current flow (black: 1 nA/nm), moderately edge-enhanced current flow

(blue: 2 nA/nm at the edges), and strongly edge-enhanced current flow (purple: 4 nA/nm at the edges) (c) Critical current density distribution $J_c(x)$ extracted from $I_{c+}(B)$ in (a), where $x$ denotes the coordinate along the junction width. Solid lines are guides to the eye. (d) Magnetic field dependence of $I_{c+}$ (symbols) for J1, compared with numerical calculations of $I_c(B)$ based on the $J_c(x)$ profile shown in the inset (solid line). The parameters $w = 317$ nm and $L_{\text{eff}} = 258$ nm are used for the calculation. Inset: Simplified model of a non-uniform $J_c(x)$ distribution.

**Fig 1.**

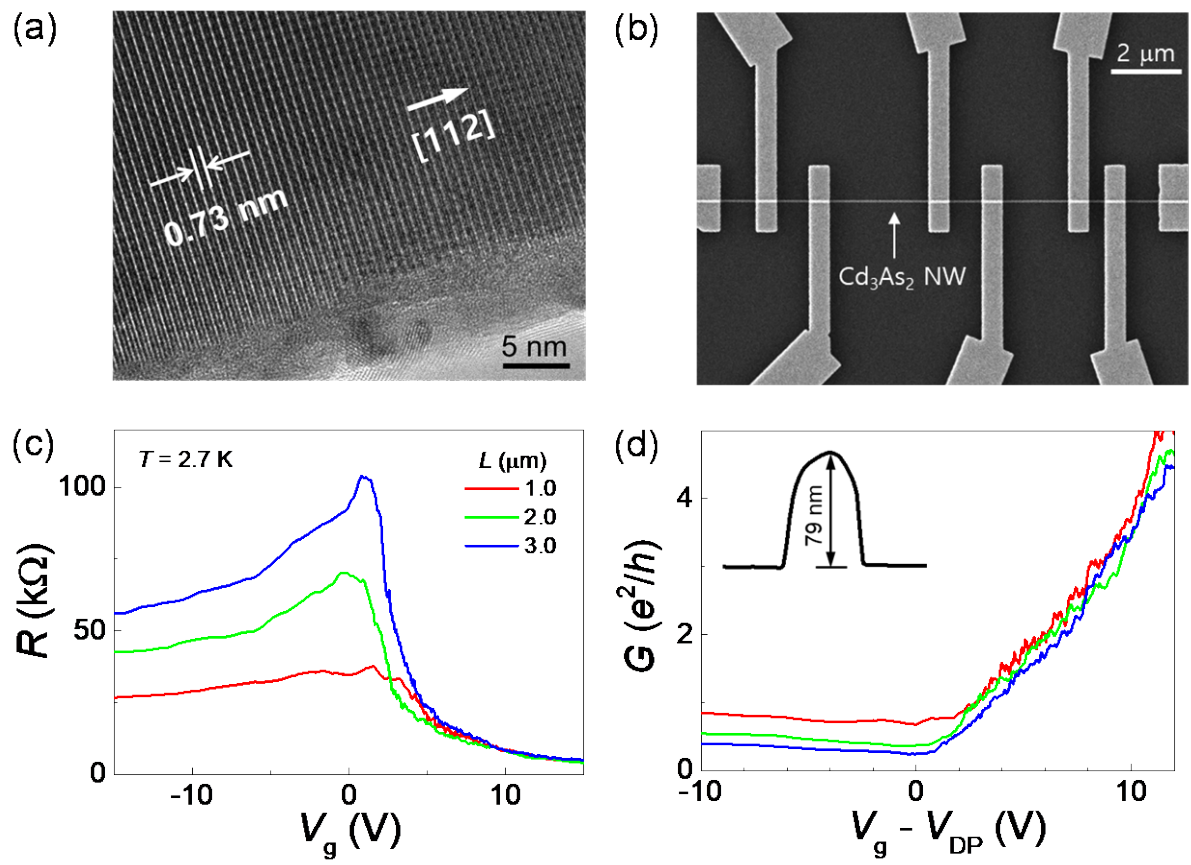

**Fig 2.**

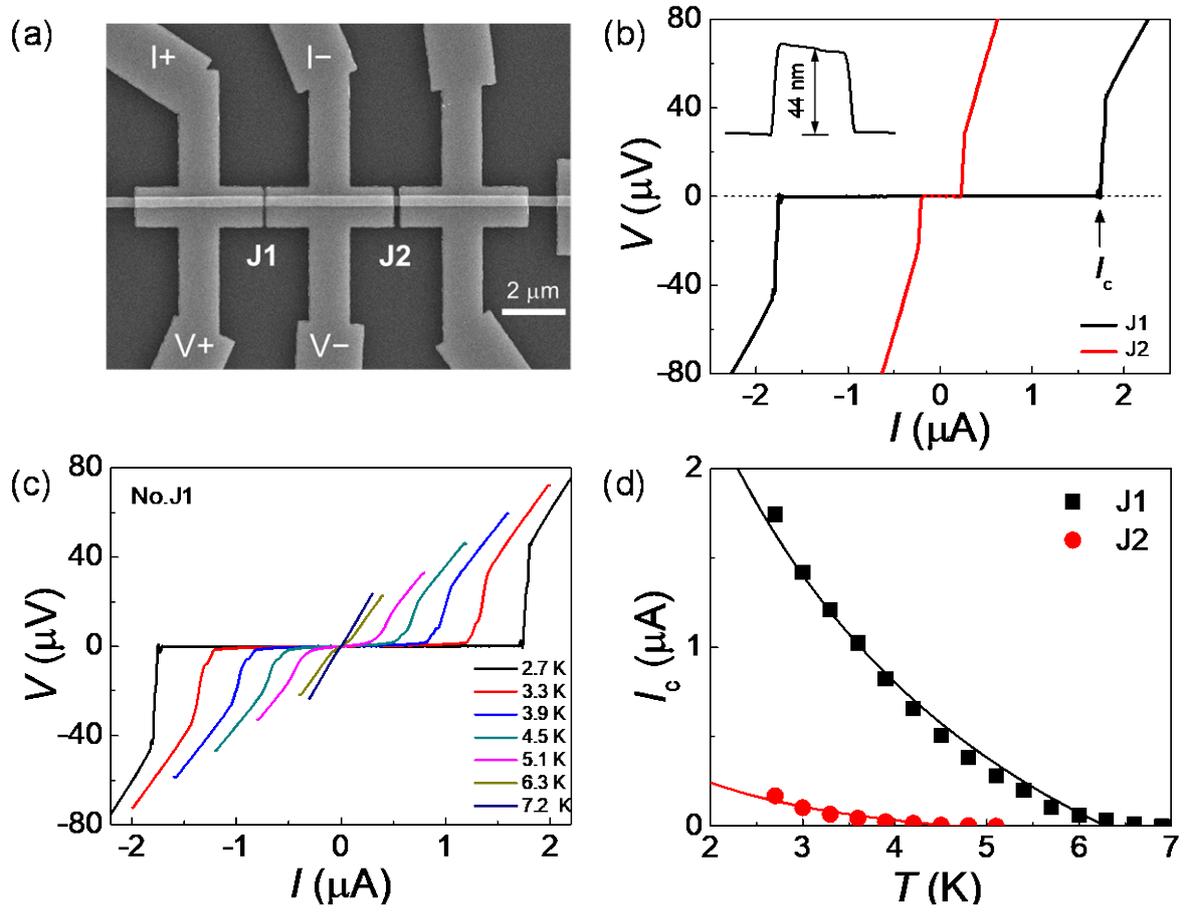

**Fig 3.**

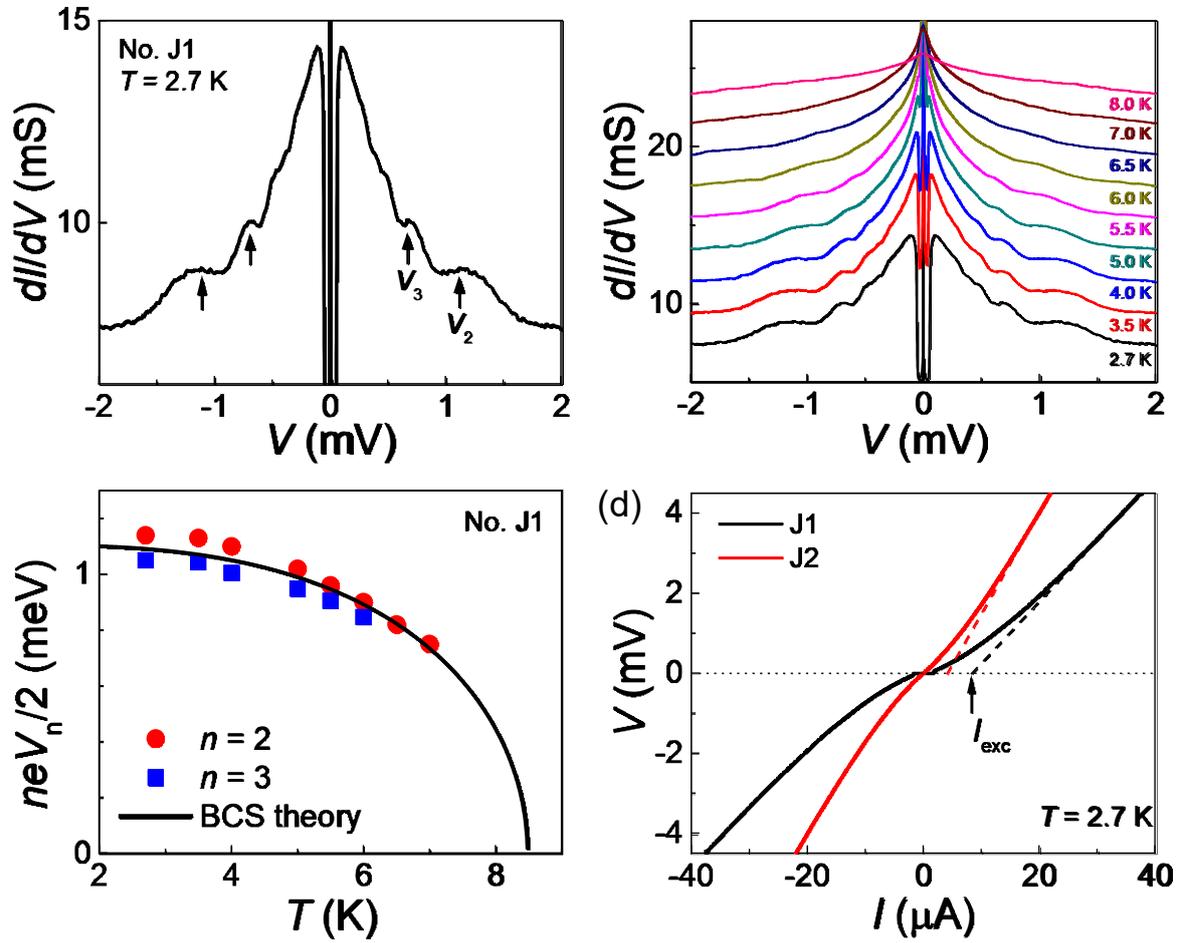

**Fig 4.**

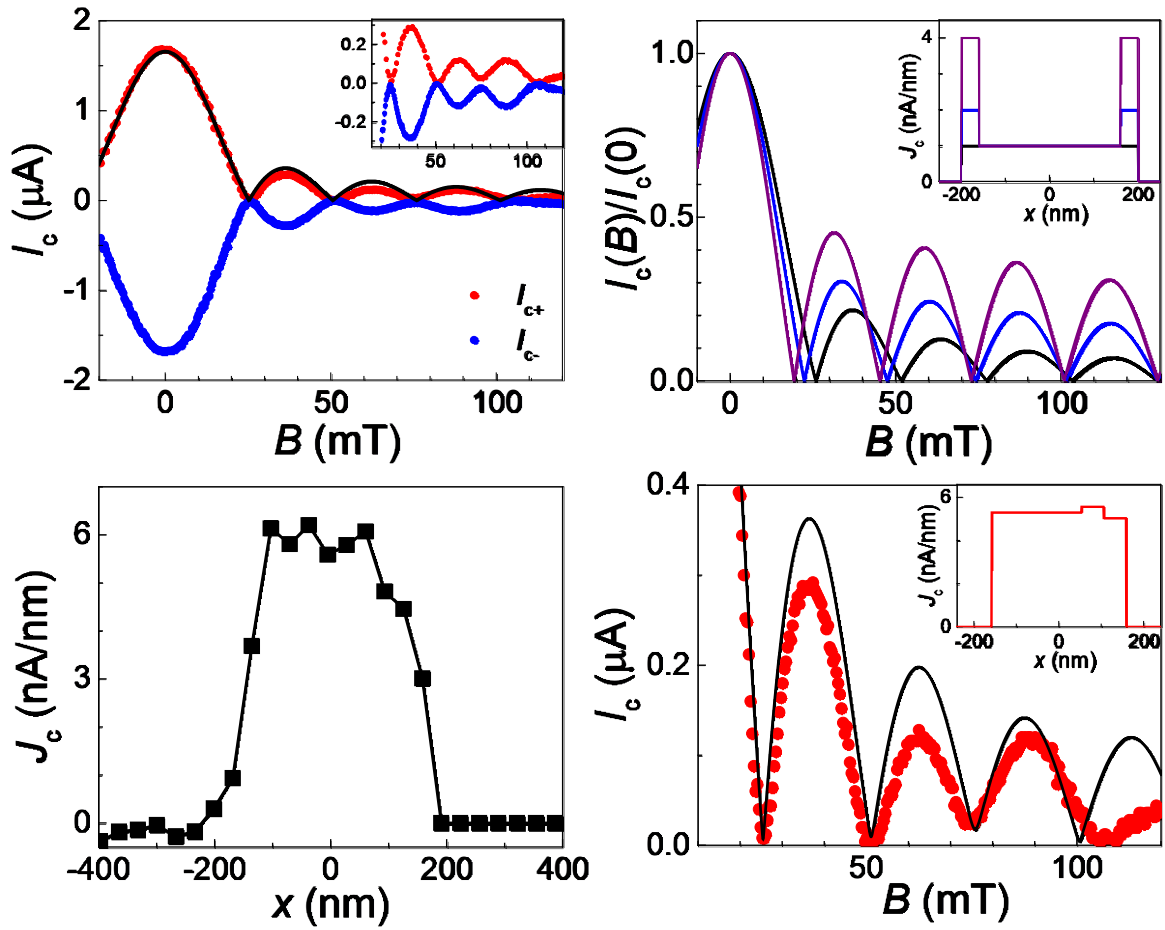